# Interference Management in 5G Reverse TDD HetNets with Wireless Backhaul: A Large System Analysis

Luca Sanguinetti, *Member, IEEE*, Aris L. Moustakas, *Senior Member, IEEE*, and Mérouane Debbah, *Fellow, IEEE*

*Abstract*—This work analyzes a heterogeneous network (HetNet), which comprises a macro base station (BS) equipped with a large number of antennas and an overlaid dense tier of small cell access points (SCAs) using a wireless backhaul for data traffic. The static and low mobility user equipment terminals (UEs) are associated with the SCAs while those with medium-to-high mobility are served by the macro BS. A reverse time division duplexing (TDD) protocol is used by the two tiers, which allows the BS to locally estimate both the intra-tier and inter-tier channels. This knowledge is then used at the BS either in the uplink (UL) or in the downlink (DL) to simultaneously serve the macro UEs (MUEs) and to provide the wireless backhaul to SCAs. A concatenated linear precoding technique employing either zero-forcing (ZF) or regularized ZF is used at the BS to simultaneously serve MUEs and SCAs in DL while nulling interference toward those SCAs in UL. We evaluate and characterize the performance of the system through the power consumption of UL and DL transmissions under the assumption that target rates must be satisfied and imperfect channel state information is available for MUEs. The analysis is conducted in the asymptotic regime where the number of BS antennas and the network size (MUEs and SCAs) grow large with fixed ratios. Results from large system analysis are used to provide concise formulae for the asymptotic UL and DL transmit powers and precoding vectors under the above assumptions. Numerical results are used to validate the analysis in different settings and to make comparisons with alternative network architectures.

## I. Introduction

The biggest challenge for next generation wireless communication systems (5G) today is to support the ever-growing demands for higher date rates and to ensure a consistent quality of service (QoS) throughout the entire network [1]. Meeting these demands requires to increase network capacity by a factor of a thousand over the next years [2]. At the same time, the power consumption of the information and communication technology industry and the corresponding energy-related pollution are becoming major societal and economical concerns [3]. Hence, more cellular network capacity on the one hand and less energy consumption on the other are seemingly contradictory future requirements on 5G. Since spectral resources are scarce, there is a broad consensus that this can only be achieved with a substantial network densification. In general, there are two different approaches for this, namely, large-scale or "massive" MIMO systems [4], [5] and small-cell networks [6]. The first approach relies on using arrays with a few hundred antennas simultaneously serving many tens of user equipment terminals (UEs) in the same frequency-time resource. The basic premise behind massive MIMO is to reap all the benefits of conventional MIMO, but on a much greater scale [5]. The second approach relies on a very dense deployment of low-cost and low-power small-cell access points (SCAs) possibly equipped with cognitive and co-operative functionalities. Although promising, each technology alone is unlikely to meet the QoS and capacity requirements for 5G [7]. On the other hand, a promising solution is a two-tier heterogeneous network (HetNet) in which the two above technologies coexist and interplay with each other in order to improve network performance [1]. In particular, massive MIMO is used to ensure outdoor coverage and to serve mobile UEs (allowing for handoff minimization), while SCAs act as the main capacity-driver for indoor and outdoor UEs with low mobility. While conventional base stations (BSs) are typically connected through a high capacity wired backhaul network, the same is not true for SCAs, which are likely to be connected via an unreliable backhaul infrastructure whose features may strongly vary from case to case, with variable characteristics of error rate, delay, capacity and especially deployment cost. For such systems, the backhaul represents one of the major bottlenecks [6]. A more economical and viable alternative is to make use of the wireless link as a backhaul [8].

### A. Main contributions

In this work, we characterize and analyze the power consumption of an HetNet consisting of a massive MIMO macro tier overlaid with a second tier of SCAs. The UEs are endowed with a single antenna and have different speeds. Those associated with the SCAs are primarily static or have low mobility while the medium-to-high mobility ones are served by the macro BS. The excess antennas at the BS are used to serve the macro UEs (MUEs) and at the same time to play the role of wireless backhaul to the SCAs. The latter are divided in two groups such that the distance between SCAs belonging to the same group is maximized and the arising interference is controlled. A similar division is performed on the MUEs on the basis of their proximity to the SCAs (see Fig. 1 of Section II). On the other hand, the interference between the macro and second tier (the so-called two-tier interference) is handled using a reverse time-division-duplexing (TDD) mode, i.e., the BS is in downlink (DL) mode when the SCAs

L. Sanguinetti is with the University of Pisa, Dipartimento di Ingegneria dell'Informazione, Pisa, Italy (luca.sanguinetti@iet.unipi.it) and also with LANEAS, CentralSupélec, Gif-sur-Yvette, France (luca.sanguinetti@centralesupelec.fr). A. L. Moustakas is with Department of Physics, National & Capodistrian University of Athens, Athens, Greece (arislm@phys.uoa.gr). M. Debbah is with LANEAS, CentralSupélec, Gif-sur-Yvette, France (merouane.debbah@centralesupelec.fr) and also with the Mathematical and Algorithmic Sciences Lab, Huawei R&D, Paris, France (merouane.debbah@huawei.com).

operate in uplink (UL), and vice versa. The TDD protocol results in a channel reciprocity that enables not only the estimation of large-dimensional channels at the BS, but also an implicit coordination between the two tiers without the need of exchanging channel state information (CSI) through the wireless backhaul. A minimum-mean-square-error (MMSE) receiver is used in UL at the BS for interference mitigation. On the other hand, a concatenated linear precoding technique employing either zero-forcing (ZF) or regularized ZF (RZF) is used in DL to satisfy rate constraints and to null interference towards SCAs, thereby providing the static small cell UEs (SUEs) a high-quality UL connection with very small power. The design and analysis of the network is performed under the assumption of imperfect CSI for the MUEs (due to their mobility) and is conducted in the asymptotic regime where the number of BS antennas $N$ and the network size (MUEs and SCAs) grow large with fixed ratio.

As we shall see, the use of BS antennas for MMSE reception and precoding allows to keep the UL and DL transmit powers of all network devices at a relatively low level for small to moderate estimation errors of MUE channels. However, we show that for a given set of target rates there is a critical value of imperfect CSI beyond which the network operation becomes infeasible as it is manifested by the divergence of all powers. In this case, MUEs with high mobility have to lower their own target rates or they have to be served using other transmission protocols that dispense from CSI (such as for example space-time coding). This might also result into a substantial reduction of the served rates.

In summary, the main contributions of this work account for: *i)* the development of a reverse TDD protocol for the coexistence of a massive MIMO macro tier and a dense tier of SCAs using a wireless backhaul for data traffic; *ii)* the asymptotically design of a concatenated RZF precoding technique for meeting rate constraints under imperfect CSI of MUEs; *iii)* the large system analysis of the power consumption in the UL and DL of each tier.

*B. Comparison with related literature*

The main literature for the system under investigation and the proposed TDD protocol is represented by [9] wherein the authors propose a similar protocol to exploit the excess antennas at the BS for intra- and inter-tier interference reduction. In contrast to [9], a wireless backhaul is introduced here for the secondary tier and imperfect CSI is assumed for MUEs. The wireless backhaul forces us to modify the transmission protocol in [9] so as to account for reverse TDD not only between tiers but also between SCAs. Moreover, we are interested in evaluating the power consumption of the network rather than the average sum rate and conduct the analysis in the large system regime.

The wireless backhaul has also been recently considered in [8] and [10]. In [8], the authors focus on the scalability properties of a wireless backhaul network modelled as a random multi-antenna extended network. In [10], a two-tier network is considered under the assumption that SCAs are full-duplex devices equipped with interference cancellation capabilities. A different line of research for wireless backhaul is in the context of mm-Wave communications. In [11], for example, the use of outdoor mm-Wave communications for backhaul networking is considered and a wind sway analysis is presented to establish a notion of beam coherence time. This highlights a previously unexplored tradeoff between array size and wind-induced movement.

The impact of imperfect CSI has been investigated in [12] wherein the authors consider the DL of a multi-cell MIMO system serving UEs with large disparities in mobility. The analysis is conducted in the asymptotic regime and shows that the mobility of a UE has a detrimental effect on its own achievable rate, but has no direct impact on the other UEs. Instead, we consider a two-tier network and evaluate the impact of imperfect CSI on the power consumption in the UL and DL of each tier, while guaranteeing requested rates. Moreover, our analysis shows that orthogonal transmission resources should be allocated to highly mobile MUEs. A similar result has been pointed out in [13] and [14].

The asymptotically optimal design of linear precoding techniques has received great attention in the last years. Some results in this context can be found in [15]–[17]. In contrast to [15], [16], this work considers a two tier network and focuses on analyzing the dual problem, namely, the power consumption of the overall network subject to target rates. On the other hand, the major differences with respect to [17] are the system under investigation and the imperfect CSI assumption at the BS.

*C. Organization*

The remainder of this paper is organized as follows.[1] Next section introduces the network architecture along with the transmission protocol and channel model. Section III focuses on the UL phase of the BS and aims at computing the power required by all transmitters taking into account the arising interference. In Section IV, we consider the DL and deal with the asymptotic analysis and design of ZF and RZF. In Section V, numerical results are used to validate the theoretical analysis and make comparisons among different network architectures. In Section VI, we discuss a possible solution to overcome the mobility issues along with that of some other practical aspects such as network design, channel correlation at the BS antennas and dynamic UL-DL TDD configurations. Finally, the major conclusions and implications are drawn in Section VII.

## II. SYSTEM MODEL

We consider a HetNet where a macro tier is augmented with a certain number of low range SCAs. Each SCA possesses a single antenna and devotes its available resources to its pre-scheduled SUE. The macro BS employs $N$ transmit antennas to serve its associated single-antenna MUEs. The MUEs are assumed to be distributed within the coverage area, while the SUEs are distributed uniformly over a circle of radius $R$ around their corresponding SCA. As shown in Fig. 1, we assume that the SCAs are divided into two groups $\mathcal{S}_\mathcal{R}$ (red colour) and $\mathcal{S}_\mathcal{B}$ (blue colour). We denote by $\mathcal{M}_\mathcal{R}$ ($\mathcal{M}_\mathcal{B}$) the set collecting MUEs that are closest to SCAs in $\mathcal{S}_\mathcal{R}$ ($\mathcal{S}_\mathcal{B}$). For notational convenience, we call $\mathcal{R} = \mathcal{M}_\mathcal{R} \cup \mathcal{S}_\mathcal{R}$ and $\mathcal{B} = \mathcal{M}_\mathcal{B} \cup \mathcal{S}_\mathcal{B}$.

---

[1] The following notation is used throughout the paper. Matrices and vectors are denoted by bold letters. The superscript $\dagger$ denotes hermitian operation and $|\mathcal{S}|$ is used to denote the cardinality of the enclosed set $\mathcal{S}$. We let $\mathbf{I}_K$ denote the $K \times K$ identity matrix and use $\mathcal{CN}(\cdot, \cdot)$ to denote a multi-variate circularly-symmetric complex Gaussian distribution whereas $\mathcal{N}(\cdot, \cdot)$ stands for a real one. The notation $\xrightarrow{a.s.}$ stands for almost surely equivalent.

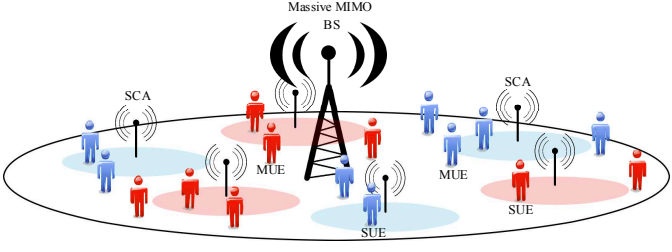

Fig. 1. Network architecture. The SCAs are divided in two different groups, namely, $\mathcal{S}_\mathcal{B}$ (blue colour) and $\mathcal{S}_\mathcal{R}$ (red colour). The same division is performed on the MUEs on the basis of their minimum distance from the SCAs.

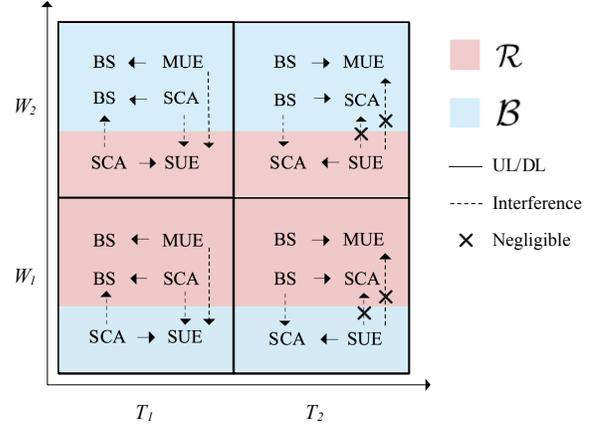

Fig. 2. Illustration of the transmission protocol. The exchange of information within each tier takes place in a reverse order, i.e., the BS is in the DL mode (BS → MUE) when the SCAs operate in the UL (SCA ← SUE), and vice versa.

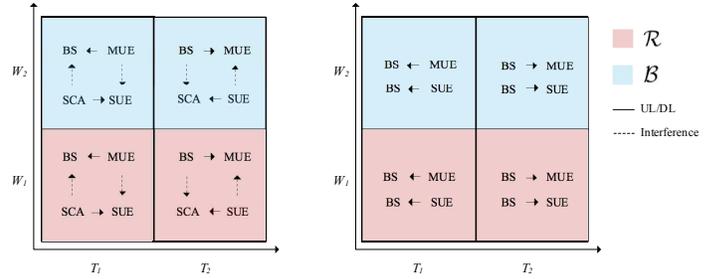

Fig. 3. Illustration of alternative transmission models and network configurations. *a)* HetNet with wired backhaul [9]; *b)* Massive MIMO. All UEs are served by the macro BS.

While conventional systems have large disparity between peak and average rates, we aim at designing the system so as to guarantee target rates or, equivalently, signal-to-interference-plus-noise ratio (SINR) values. The analysis is conducted in the asymptotic regime in which the number of BS antennas increases as the network size becomes large. A known problem with the asymptotic analysis is that the target rates are not guaranteed to be achieved when $N$ is finite and relatively small (see for example [18]). This is because the approximation errors are translated into fluctuations in the resulting SINR values. However, these errors vanish rapidly when $N$ takes large yet finite values as it is envisioned for massive MIMO systems [19], [20].

### A. Transmission protocol

The operating protocol is sketched in Fig. 2. In the frequency-time slot $(W_1, T_1)$, the MUEs and SCAs in $\mathcal{R}$ use the frequency band $W_1$ for UL transmissions (BS ← MUE and BS ← SCA) for a time interval of length $T_1$ whereas the SCAs in $\mathcal{B}$ transmit to their associated SUEs in the DL (SCA → SUE). In $(W_1, T_2)$, the reverse takes place, i.e., the BS makes use of $W_1$ for a time interval of length $T_2$ to transmit in the DL to the MUEs and SCAs in $\mathcal{R}$ whereas the SUEs associated to the SCAs in $\mathcal{B}$ use $W_1$ for UL transmissions. The frequency-time slots $(W_2, T_1)$ and $(W_2, T_2)$ are used in the dual way. As seen, the exchange of information within each tier takes place in a reverse order, i.e., the BS is in the DL mode (BS → MUE) when the SCAs operate in the UL (SCA ← SUE), and vice versa. We assume that transmissions across tiers are perfectly synchronized (the impact of asynchronous transmissions will be discussed in Section VI) and that the channel frequency response is flat over each frequency band. We also assume that $T_1 + T_2$ is upper bounded by the coherence time of the channel. In these circumstances, UL and DL channels can be considered as reciprocal and the BS can make use of UL estimates for DL transmissions (more details on this will be given later on).

**Remark 1.** *Observe that the transmission protocol described above is mainly driven by the need:* i) *to provide a wireless backhaul to the SCAs while serving MUEs and SUEs;* ii) *to propose a viable solution to counteract the arising interference. This is achieved by geographically separating co-channel SUEs and by letting the channel reciprocity condition (within each tier and between tiers) hold in order to properly exploit knowledge of the channel for precoding and decoding (more details on the channel acquisition will be given later on when required). Alternative solutions can in principle be found. In the simulations, comparisons will be made with the two transmission protocols shown in Fig. 3. In particular, the one on the left applies to a reverse TDD HetNet in which SCAs are connected to the BS through a wired backhaul [9]. On the other hand, the protocol on the right is for a massive MIMO system in which only the macro-tier is present [5].*

### B. Channel Model and Assumptions

We denote $\mathbf{h}_i^{(\mathcal{M}_\mathcal{R})} \in \mathbb{C}^{N \times 1}$ the vector whose entry $h_i^{(\mathcal{M}_\mathcal{R})}(n)$ accounts for the instantaneous propagation channel between the $i$th MUE in $\mathcal{M}_\mathcal{R}$ and the $n$th antenna at the BS. For mathematical convenience, we assume that the BS antennas are uncorrelated (see Section VI-C for a discussion on this assumption). In these circumstances, the channel vector $\mathbf{h}_i^{(\mathcal{M}_\mathcal{R})}$ can be modelled as [16], [21]:

$$\mathbf{h}_i^{(\mathcal{M}_\mathcal{R})} = \sqrt{Nl(\mathbf{x}_i)}\mathbf{z}_i \qquad (1)$$

where $\mathbf{x}_i$ denotes the position of MUE $i$ in $\mathcal{M}_\mathcal{R}$ (computed with respect to the BS), $\mathbf{z}_i \sim \mathcal{CN}(0, \mathbf{I}_N N^{-1})$ accounts for the small-scale fading channel and $l(\mathbf{x}_i) : \mathbb{R}^2 \to \mathbb{R}^+$ is the average channel gain due to pathloss at distance $\|\mathbf{x}_i\|$. Since the forthcoming analysis does not depend on a particular choice of $l(\mathbf{x}_i)$ as long as it is a decreasing function of the distance $\|\mathbf{x}_i\|$ and is bounded from below, we keep it generic [22]. Accordingly, we let $\mathbf{H}^{(\mathcal{M}_\mathcal{R})} = [\mathbf{h}_1^{(\mathcal{M}_\mathcal{R})} \mathbf{h}_2^{(\mathcal{M}_\mathcal{R})} \cdots \mathbf{h}_{|\mathcal{M}_\mathcal{R}|}^{(\mathcal{M}_\mathcal{R})}] \in \mathbb{C}^{N \times |\mathcal{M}_\mathcal{R}|}$ be the matrix collecting the channels of all MUEs in $\mathcal{M}_\mathcal{R}$. The same model is adopted for the SCAs and SUEs. In particular, we let $\mathbf{H}^{(\mathcal{S}_\mathcal{R})} \in \mathbb{C}^{N \times |\mathcal{S}_\mathcal{R}|}$ and $\mathbf{H}^{(\mathcal{S}_\mathcal{B})} \in \mathbb{C}^{N \times |\mathcal{S}_\mathcal{B}|}$ be the matrices collecting the channel gains from the BS antennas and the SCAs in $\mathcal{S}_\mathcal{R}$ and $\mathcal{S}_\mathcal{B}$, respectively.

In all subsequent discussions, we assume that only an estimate $\widehat{\mathbf{H}}^{(\mathcal{M}_R)}$ of $\mathbf{H}^{(\mathcal{M}_R)}$ is available. In particular, we model each vector $\widehat{\mathbf{h}}_i^{(\mathcal{M}_R)}$ of $\widehat{\mathbf{H}}^{(\mathcal{M}_R)}$ as [16]

$$\widehat{\mathbf{h}}_i^{(\mathcal{M}_\mathcal{R})} = \sqrt{N l(\mathbf{x}_i)} \left( \sqrt{1 - \tau_i^2} \mathbf{z}_i + \tau_i \mathbf{v}_i \right) = \sqrt{N l(\mathbf{x}_i)} \widehat{\mathbf{z}}_i \quad (2)$$

where $\mathbf{v}_i \sim \mathcal{CN}(0, 1/N \mathbf{I}_N)$ accounts for the independent channel estimation errors. The parameter $\tau_i \in [0,1]$ reflects the accuracy or quality of the channel estimate, i.e., $\tau_i = 0$ corresponds to perfect CSI, whereas for $\tau_i = 1$ the CSI is completely uncorrelated to the true channel.

Observe that imperfect CSI arises naturally for MUEs as a consequence of mobility [12], [16]. Since SCAs occupy fixed positions in the network, then the propagation channels remain constant for a sufficiently large number of phases to be accurately estimated. For this reason, in all subsequent discussions we assume that $\mathbf{H}^{(\mathcal{S}_\mathcal{R})}$ and $\mathbf{H}^{(\mathcal{S}_\mathcal{B})}$ are perfectly known at the BS (i.e., $\tau_i = 0$ if $i \in S_\mathcal{R}$ or $S_\mathcal{B}$). The same assumption is made for the SUE channels.

### III. LARGE SYSTEM ANALYSIS OF THE MACRO-TIER INTERFERENCE IN UL

We start dealing with the case in which the BS is in UL mode. Without loss of generality, the frequency-time slot $(W_1, T_1)$ of Fig. 2 is considered.[2] As seen, two instances of interference appear. One comes from UL signals of MUEs and SCAs in $\mathcal{R}$ and affects the receiving SUEs in $\mathcal{B}$ whereas the other accounts for the interference that the BS experiences from the DL mode of SCAs in $\mathcal{S}_\mathcal{B}$. The former interference is limited due to the geographical separation of co-channel SUEs. Although inherently mitigated by the geographical separation of $\mathcal{R}$ and $\mathcal{B}$, this interference can be a limiting factor due to the potentially large number of transmitting MUEs (see also the analysis in Section VI-B) and the lack of spatial degrees of freedom at the SUEs. Therefore, it cannot be neglected but it must be taken into account while designing the transmit powers of MUEs (in UL) and SCAs (in DL). With regards to the latter interference, it easily turns out that the large number of antennas provides an effective means to mitigate its detrimental effects and at the same time to simultaneously serve all transmitters in $\mathcal{R}$. For this purpose, we assume that an MMSE receiver is employed at the BS.

[2]The same analysis can be performed for $(W_2, T_1)$.

For notational convenience, we denote $K = |\mathcal{R}|$ the total number of transmitters (MUEs and SCAs) in $\mathcal{R}$ and call $S = |\mathcal{S}_\mathcal{B}|$ the number of SCAs in $\mathcal{S}_\mathcal{B}$. We also let

$$c = \frac{K}{N} \quad \text{and} \quad c_S = \frac{S}{N}. \quad (3)$$

We denote $\mathbf{H} = [\mathbf{h}_1, \mathbf{h}_2, \ldots, \mathbf{h}_K] = [\mathbf{H}^{(\mathcal{M}_\mathcal{R})}, \mathbf{H}^{(\mathcal{S}_\mathcal{R})}] \in \mathbb{C}^{N \times K}$ the matrix collecting the instantaneous UL channels of MUEs and SCAs in $\mathcal{R}$ and denote $\{p_k^{(\mathcal{R}, \mathrm{ul})}\}$ the corresponding UL transmit powers. Letting $\mathbf{G} = [\mathbf{g}_1, \mathbf{g}_2, \ldots, \mathbf{g}_K] \in \mathbb{C}^{N \times K}$ be the MMSE matrix as given in (4) (shown at the top of next page), the UL achievable rate for device $k$ in $\mathcal{R}$ is [21]

$$R_k^{(\mathcal{R}, \mathrm{ul})} = \log_2 \left( 1 + \mathrm{SINR}_k^{(\mathcal{R}, \mathrm{ul})} \right) \quad (6)$$

with $\mathrm{SINR}_k^{(\mathcal{R}, \mathrm{ul})}$ given by (5) (shown at the top of next page) where $\sigma^2$ accounts for thermal noise and $p_s^{(\mathcal{S}_\mathcal{B}, \mathrm{dl})}$ is the DL transmit power of SCA $s$ in $\mathcal{S}_\mathcal{B}$. As mentioned earlier, we assume that the MMSE receiver operates under the assumption of imperfect knowledge of $\mathbf{H}^{(\mathcal{M}_\mathcal{R})}$. This amounts to setting $\mathbf{G}$ as in (4) where $\widehat{\mathbf{h}}_k$ is the $k$th column of $\widehat{\mathbf{H}}$ given by $\widehat{\mathbf{H}} = [\widehat{\mathbf{H}}^{(\mathcal{M}_\mathcal{R})} \mathbf{H}^{(\mathcal{S}_\mathcal{R})}]$. Observe that $\{p_k^{(\mathcal{R}, \mathrm{ul})}\}$ and $\{p_s^{(\mathcal{S}_\mathcal{B}, \mathrm{dl})}\}$ are assumed to be perfectly known at the BS. This information can be easily acquired through signalling [23].

**Remark 2.** *It is worth observing that in the frequency-time slot $(W_1, T_1)$ under consideration $\mathbf{H}^{(\mathcal{S}_\mathcal{R})}$ can be easily acquired at the BS using UL pilots from SCAs in $\mathcal{S}_\mathcal{R}$. On the other hand, the estimation of $\mathbf{H}^{(\mathcal{S}_\mathcal{B})}$ must be performed in a different way since the UL mode (BS $\leftarrow$ SCA) for $\mathcal{S}_\mathcal{B}$ takes place over the frequency band $W_2$. A possible solution might consist in using pilots that the SCAs in $\mathcal{S}_\mathcal{B}$ send in DL (SCA $\to$ SUE) to their associated SUEs. An alternative approach might be to periodically switch the operations of frequency bands $W_1$ and $W_2$ [24].*

The aim of this section is to compute the UL and DL transmit powers $\{p_k^{(\mathcal{R}, \mathrm{ul})}\}$ and $\{p_s^{(\mathcal{S}_\mathcal{B}, \mathrm{dl})}\}$ required to meet target requirements $\{r_k^{(\mathcal{R}, \mathrm{ul})}\}$ and $\{r_s^{(\mathcal{S}_\mathcal{B}, \mathrm{dl})}\}$ under imperfect CSI of MUEs. For notational convenience, the superscripts $^{(\mathrm{ul})}$ and $^{(\mathrm{dl})}$ are dropped in the sequel. We start assuming that the downlink powers $\{p_s^{(\mathcal{S}_\mathcal{B})}\}$ are fixed and given. Let $\gamma_k = 2^{r_k^{(\mathcal{R})}} - 1$ be the target SINR value for user $k$ in $\mathcal{R}$. Then, the following lemma can be proved.

**Lemma 1.** *If the MMSE receiver in (4) is employed at the BS, then in the limit $N, K, S \to \infty$ with $c + c_S \in (0,1)$ we have that $p_k^{(\mathcal{R})} - \overline{p}_k^{(\mathcal{R})} \xrightarrow{a.s.} 0$ with*

$$\overline{p}_k^{(\mathcal{R})} = \frac{1}{\xi \delta} \frac{\gamma_k}{l(\mathbf{x}_k)(1 - \tau_k^2)} \quad (7)$$

*with $\tau_k = 0$ if $k \in \mathcal{S}_\mathcal{R}$. The quantities $\xi$ and $\delta$ are computed as the unique solutions to (8) and (9) (shown at the top of next page).*

*Proof:* The proof relies on using the same random matrix theory results of [16] to obtain the deterministic equivalent of the SINR in (5) for a given set of $\{p_s^{(\mathcal{S}_\mathcal{B})}\}$. This result is then used to compute the deterministic equivalents of the powers $\{p_k^{(\mathcal{R})}\}$ that are required in the asymptotic regime to achieve

$$\mathbf{G} = \left( \sum_{i=1}^{K} p_i^{(\mathcal{R},\mathrm{ul})} \widehat{\mathbf{h}}_k \widehat{\mathbf{h}}_k^{\dagger} + \sum_{s=1}^{S} p_s^{(\mathcal{S}_\mathcal{B},\mathrm{dl})} \mathbf{h}_s^{(\mathcal{S}_\mathcal{B})} \mathbf{h}_s^{(\mathcal{S}_\mathcal{B})\dagger} + N\sigma^2 \mathbf{I}_N \right)^{-1} \widehat{\mathbf{H}} \quad (4)$$

$$\mathrm{SINR}_k^{(\mathcal{R},\mathrm{ul})} = \frac{p_k^{(\mathcal{R},\mathrm{ul})} \left|\mathbf{g}_k^{\dagger} \mathbf{h}_k\right|^2}{\sum_{i=1,i\neq k}^{K} p_i^{(\mathcal{R},\mathrm{ul})} \left|\mathbf{g}_k^{\dagger} \mathbf{h}_i\right|^2 + \sum_{s=1}^{S} p_s^{(\mathcal{S}_\mathcal{B},\mathrm{dl})} \left|\mathbf{g}_k^{\dagger} \mathbf{h}_s^{(\mathcal{S}_\mathcal{B})}\right|^2 + \sigma^2 \|\mathbf{g}_k\|^2} \quad (5)$$

$$\xi = \frac{1}{\sigma^2}\left[1 - \frac{1}{N}\sum_{i=1}^{K}\frac{\gamma_i}{\delta(1-\tau_i^2)+\gamma_i} - \frac{1}{N}\sum_{s=1}^{S}\frac{p_s^{(\mathcal{S}_\mathcal{B})} l(\mathbf{x}_s)\xi}{1+p_s^{(\mathcal{S}_\mathcal{B})} l(\mathbf{x}_s)\xi}\right] \quad (8)$$

$$\delta = \frac{\frac{1}{N}\sum_{i=1}^{K}\frac{\gamma_i}{\delta}\frac{(1-\tau_i^2)}{\left(\delta(1-\tau_i^2)+\frac{\gamma_i}{\delta}\right)^2} + \frac{1}{N}\sum_{s=1}^{S}\frac{p_s^{(\mathcal{S}_\mathcal{B})} l(\mathbf{x}_s)\xi}{\left(1+p_s^{(\mathcal{S}_\mathcal{B})} l(\mathbf{x}_s)\xi\right)^2} + \xi\sigma^2}{\frac{1}{N}\sum_{i=1}^{K}\frac{\gamma_i}{\delta}\frac{\tau_i^2}{1-\tau_i^2} + \frac{1}{N}\sum_{i=1}^{K}\frac{\gamma_i}{\delta}\frac{\gamma_i(1-\tau_i^2)^2}{\left((1-\tau_i^2)+\frac{\gamma_i}{\delta}\right)^2} + \frac{1}{N}\sum_{s=1}^{S}\frac{p_s^{(\mathcal{S}_\mathcal{B})} l(\mathbf{x}_s)\xi}{\left(1+p_s^{(\mathcal{S}_\mathcal{B})} l(\mathbf{x}_s)\xi\right)^2} + \xi\sigma^2}. \quad (9)$$

---

the SINR constraints $\{\gamma_k\}$. The sketch of the proof is given in the Appendix. ∎

The evaluation of $\{p_s^{(\mathcal{S}_\mathcal{B})}\}$ requires the computation of the DL SINR of the $s$th SUE in $\mathcal{S}_\mathcal{B}$, which is given by

$$\mathrm{SINR}_s^{(\mathcal{S}_\mathcal{B})} = \frac{p_s^{(\mathcal{S}_\mathcal{B})}|h_s|^2}{\sigma^2 + \sum_{k=1}^{K} p_k^{(\mathcal{R})}|h_{s,k}|^2} \quad (10)$$

where $h_s$ is the channel propagation coefficient from its serving SCA whereas $h_{s,k}$ is the channel coefficient of the $k$th interfering UL transmission in $\mathcal{R}$.[3] Observe that if $K$ is large, then the interference term in (10) can be reasonably assumed to be deterministic and equal to its mean. More specifically, under the assumption that all powers $p_k^{(\mathcal{R})}$ are finite and the cell size is fixed, using the law of large numbers yields

$$\frac{1}{K}\sum_{k=1}^{K} p_k^{(\mathcal{R})}|h_{s,k}|^2 \xrightarrow[K\to\infty]{} \frac{1}{K}\sum_{k=1}^{K} p_k^{(\mathcal{R})} l(\mathbf{x}_{s,k}) \quad (11)$$

where $\mathbf{x}_{s,k}$ denotes the distance of SUE $s$ from transmitter $k$ in $\mathcal{R}$. In contrast, the power of the useful signal in (10) is a random quantity that depends on the fluctuations of $|h_s|^2$. To overcome this issue, we resort to the ergodic mutual information as a metric[4]. Using (11), we find the following asymptotic result:

$$\mathbb{E}_{h_s}\left\{\log_2\left(1+\mathrm{SINR}_s^{(\mathcal{S}_\mathcal{B})}\right)\right\} \xrightarrow[K\to\infty]{} \frac{e^{1/\overline{\mathrm{SINR}}_s^{(\mathcal{S}_\mathcal{B})}}}{\log(2)} \mathrm{E}_1\left(1/\overline{\mathrm{SINR}}_s^{(\mathcal{S}_\mathcal{B})}\right) \quad (12)$$

where $\mathrm{E}_1(z) = \int_z^\infty dt \frac{e^{-t}}{t}$ is the exponential integral of order 1 whereas

$$\overline{\mathrm{SINR}}_s^{(\mathcal{S}_\mathcal{B})} = \frac{p_s^{(\mathcal{S}_\mathcal{B})} l(\mathbf{x}_s)}{\sigma^2 + \sum_{k=1}^{K} \bar{p}_k^{(\mathcal{R})} l(\mathbf{x}_{s,k})} \quad (13)$$

---

[3]In writing (10), we have not neglected the interference coming from the other SCAs in $\mathcal{S}_\mathcal{B}$ as they are assumed to be relatively spaced apart.

[4]Observe that an alternative route might be that of using the outage capacity criterion.

---

with $\bar{p}_k^{(\mathcal{R})}$ being obtained from (7). Imposing (12) equal to $r_s^{(\mathcal{S}_\mathcal{B})}$ and inverting the exponential integral provides the target SINR $\gamma_s$. Setting $\overline{\mathrm{SINR}}_s^{(\mathcal{S}_\mathcal{B})} = \gamma_s$ and using (7), the DL power of SCA $s$ satisfying the target rate constraint is obtained as

$$p_s^{(\mathcal{S}_\mathcal{B})} = \frac{\gamma_s}{l(\mathbf{x}_s)}\left(\sigma^2 + \frac{1}{\xi\delta}\sum_{k=1}^{K}\frac{\gamma_k}{1-\tau_k^2}\frac{l(\mathbf{x}_{k,s})}{l(\mathbf{x}_k)}\right) \quad (14)$$

with $\tau_k = 0$ if $k \in \mathcal{S}_\mathcal{R}$ and $\gamma_k = 2^{r_k^{(\mathcal{R})}} - 1$. Plugging the above result into (8) and (9), it follows that the computation of the powers $\{p_k^{(\mathcal{R})}\}$ and $\{p_s^{(\mathcal{S}_\mathcal{B})}\}$ reduces to the easy task of finding $\xi$ and $\delta$ as the unique solutions of (8) and (9) that depend only on system parameters (such as imperfect CSI factors $\{\tau_k\}$, SINR constraints $\{\gamma_k\}$ and number of MUEs and SCAs).

From the above results, it follows that the imperfect CSI coefficients $\{\tau_k\}$ impact both $\xi$ and $\delta$ in (8) and (9). In particular, from (7) it follows that $(\xi\delta)^{-1}$ can be thought of as the fractional UL power increase of all transmitters (MUEs and SCAs) in $\mathcal{R}$. Interestingly, this happens even though only the MUE channels are estimated erroneously while perfect CSI is assumed for SCAs. To gain some insights on the maximum tolerable level of imperfect CSI, we now look for which values of $\{\tau_k\}$ and $\{\gamma_k\}$ the power diverges. This amounts to solving (9) for $\delta \to 0$ since $\xi$ can be shown to remain finite even when all powers diverge. In doing so, it turns out that (9) has finite positive solutions only if

$$\frac{1}{N}\sum_{k\in\mathcal{M}_\mathcal{R}} \gamma_k \frac{\tau_k^2}{1-\tau_k^2} \leq 1 - c - c_S. \quad (15)$$

If this condition is not met, then all powers diverge. If $\tau_k = \tau$ for any $k \in \mathcal{M}_\mathcal{R}$ one gets that $\tau$ has to be smaller than $\tau_{\mathrm{MMSE}}^{(\max)}$ given by

$$\tau_{\mathrm{MMSE}}^{(\max)} = \left(1 + \bar{\gamma}^{(\mathcal{M})}\frac{c}{1-c-c_S}\right)^{-1/2} \quad (16)$$

where $\bar{\gamma}^{(\mathcal{M})}$ stands for

$$\bar{\gamma}^{(\mathcal{M})} = \frac{1}{K} \sum_{k \in \mathcal{M}_\mathcal{R}} \gamma_k \qquad (17)$$

average SINR requirements of all devices in $\mathcal{M}_\mathcal{R}$.

## IV. LARGE SYSTEM ANALYSIS OF THE MACRO-TIER INTERFERENCE IN DL

We now consider the case in which the BS is in DL mode. Without loss of generality, the frequency-time slot $(W_1, T_2)$ of Fig. 2 is considered. As for the UL, two instances of interference arise. The interference experienced by MUEs and SCAs from UL transmissions in $\mathcal{S}_\mathcal{B}$ can be reasonably neglected since the number of transmitting SUEs is relatively small (one per SCA) and geographically far away from the MUEs and SCAs in $\mathcal{R}$. On the other hand, the interference from BS to the SCAs in UL must be properly mitigated to avoid a severe degradation of the network performance. For this purpose, we assume that the BS makes use of linear precoding and sacrifices some of its degrees of freedom (or excess antennas) to simultaneously serve all receivers in $\mathcal{R}$ and at the same time to null the interference towards $\mathcal{S}_\mathcal{B}$. We let $\mathbf{V} = [\mathbf{v}_1, \mathbf{v}_2, \ldots, \mathbf{v}_K] \in \mathbb{C}^{N \times K}$ be the precoding matrix and denote $p_k^{(\mathcal{R},\text{dl})}$ the DL transmit power assigned to the $k$th device in $\mathcal{R}$. The total DL transmit power at the BS is [25]

$$P^{(\mathcal{R},\text{dl})} = \sum_{k=1}^{K} p_k^{(\mathcal{R},\text{dl})} \|\mathbf{v}_k\|^2 \qquad (18)$$

whereas the achievable DL rate for a generic receiver $k$ in $\mathcal{R}$ is $R_k^{(\mathcal{R},\text{dl})} = \log_2(1 + \text{SINR}_k^{(\mathcal{R},\text{dl})})$ with

$$\text{SINR}_k^{(\mathcal{R},\text{dl})} = \frac{p_k^{(\mathcal{R},\text{dl})} \left|\mathbf{h}_k^\dagger \mathbf{v}_k\right|^2}{\sum_{i=1, i \neq k}^{K} p_i^{(\mathcal{R},\text{dl})} \left|\mathbf{h}_k^\dagger \mathbf{v}_i\right|^2 + \sigma^2}. \qquad (19)$$

We impose $R_k^{(\mathcal{R},\text{dl})} = r_k^{(\mathcal{R},\text{dl})}$ or, equivalently, $\text{SINR}_k^{(\mathcal{R},\text{dl})} = \gamma_k$ with $\gamma_k = 2^{r_k^{(\mathcal{R},\text{dl})}} - 1$. Thanks to the reciprocity of UL and DL channels, the BS can exploit UL estimates for DL transmissions. As for the UL, we assume that perfect knowledge of $\mathbf{H}^{(\mathcal{S}_\mathcal{B})}$ and $\mathbf{H}^{(\mathcal{S}_\mathcal{R})}$ is available while imperfect CSI is assumed for $\mathbf{H}^{(\mathcal{M}_\mathcal{R})}$. For notational convenience, the superscript $^{(\text{dl})}$ is dropped in the sequel.

The complete elimination of the macro-tier interference at SCAs in $\mathcal{S}_\mathcal{B}$ can be achieved by constraining the precoding matrix $\mathbf{V}$ to lie in the null space of $\mathbf{H}^{(\mathcal{S}_\mathcal{B})}$. Under the assumption of perfect knowledge of $\mathbf{H}^{(\mathcal{S}_\mathcal{B})}$, this is achieved setting $\mathbf{V} = \mathbf{T}^{(\mathcal{S}_\mathcal{B})} \mathbf{F}$ where $\mathbf{F} = [\mathbf{f}_1, \mathbf{f}_2, \ldots, \mathbf{f}_K] \in \mathbb{C}^{N \times K}$ is a design matrix and $\mathbf{T}^{(\mathcal{S}_\mathcal{B})} \in \mathbb{C}^{N \times N}$ is obtained as

$$\mathbf{T}^{(\mathcal{S}_\mathcal{B})} = \mathbf{I}_N - \mathbf{H}^{(\mathcal{S}_\mathcal{B})} \left(\mathbf{H}^{(\mathcal{S}_\mathcal{B})\dagger} \mathbf{H}^{(\mathcal{S}_\mathcal{B})}\right)^{-1} \mathbf{H}^{(\mathcal{S}_\mathcal{B})\dagger}. \qquad (20)$$

Let $\mathbf{U} = \mathbf{T}^{(\mathcal{S}_\mathcal{B})} \mathbf{H} \in \mathbb{C}^{N \times K}$ be the composite channel and denote $\widehat{\mathbf{U}}$ its corresponding estimate defined as $\widehat{\mathbf{U}} = \mathbf{T}^{(\mathcal{S}_\mathcal{B})} \widehat{\mathbf{H}}$ (under the assumptions given above) where $\widehat{\mathbf{H}} = [\widehat{\mathbf{H}}^{(\mathcal{M}_R)} \, \mathbf{H}^{(\mathcal{S}_R)}]$. The matrix $\widehat{\mathbf{U}}$ is used in the sequel to design $\mathbf{F}$ according to the RZF and ZF criteria.

### A. Regularized Zero Forcing

We start assuming that $\mathbf{F}$ takes the following form:

$$\mathbf{F} = \left(\widehat{\mathbf{U}} \mathbf{\Lambda}^{-1} \widehat{\mathbf{U}}^\dagger + N\rho \mathbf{I}_N\right)^{-1} \widehat{\mathbf{U}} \qquad (21)$$

where $\mathbf{\Lambda} = \text{diag}\{l(\mathbf{x}_1), l(\mathbf{x}_2), \ldots, l(\mathbf{x}_K)\}$ and $\rho > 0$ is a design parameter. As in [16], $\rho$ is multiplied by $N$ to ensure that $\rho$ converges to a constant as $N, K \to \infty$. We refer to the concatenated matrix $\mathbf{V}_{\text{RZF}} = \mathbf{T}^{(\mathcal{S}_\mathcal{B})} \mathbf{F}$ as RZF and denote

$$P_{\text{RZF}}^{(\mathcal{R})} = \sum_{k=1}^{K} p_k^{(\mathcal{R})} \left\|\mathbf{T}^{(\mathcal{S}_\mathcal{B})} \mathbf{f}_k\right\|^2 \qquad (22)$$

its corresponding transmit power. Observe that in the design of $\mathbf{F}$ in (21) we exploit knowledge of the average channel attenuations $\{l(\mathbf{x}_i)\}$ through $\mathbf{\Lambda}$. This information can be easily observed and estimated accurately at the BS because it changes slowly with time (relative to the small-scale fading) even for MUEs with medium-to-high mobility. This choice is inspired to [17] wherein it is proved that in the downlink of a single-tier MIMO system with perfect CSI, such a kind of RZF is asymptotically equivalent to the optimal linear precoder when the same rate constraints are imposed for all UEs. Due to the imperfect CSI and the projection into the null space of SCAs in $\mathcal{S}_\mathcal{B}$, the results in [17] do not apply to network under investigation. However, the use of $\{l(\mathbf{x}_i)\}$ is instrumental to get a closed form expression for the optimal $\rho$ in (21).

For convenience, we let

$$A = \frac{1}{K} \sum_{k \in \mathcal{M}_\mathcal{R}} \frac{\gamma_k}{(1 - \tau_k^2) l(\mathbf{x}_k)} + \frac{1}{K} \sum_{k \in \mathcal{S}_\mathcal{R}} \frac{\gamma_k}{l(\mathbf{x}_k)} \qquad (23)$$

and

$$B = \frac{1}{K} \sum_{k \in \mathcal{M}_\mathcal{R}} \gamma_k \frac{\tau_k^2}{1 - \tau_k^2} \qquad (24)$$

and denote

$$\bar{\gamma} = \frac{1}{K} \sum_{k=1}^{K} \gamma_k \qquad (25)$$

the average SINR requirement of all devices in $\mathcal{R}$.

**Lemma 2.** *If RZF is used and $N, K, S \to \infty$ with $c + c_S \in (0, 1)$, then $P_{\text{RZF}}^{(\mathcal{R})} - \overline{P}_{\text{RZF}}^{(\mathcal{R})} \xrightarrow{a.s.} 0$ where $\overline{P}_{\text{RZF}}^{(\mathcal{R})}$ is given by*

$$\overline{P}_{\text{RZF}}^{(\mathcal{R})} = c\sigma^2 \frac{A}{\rho^\star \bar{\gamma} - cB} \qquad (26)$$

*where the optimal $\rho$ is computed as*

$$\rho^\star = \frac{1 - c_S}{\bar{\gamma}} - \frac{c}{1 + \bar{\gamma}}. \qquad (27)$$

*Also, $p_k^{(\mathcal{R})} - \overline{p}_k^{(\mathcal{R})} \xrightarrow{a.s.} 0$ with*

$$\overline{p}_k^{(\mathcal{R})} = \frac{\gamma_k}{l(\mathbf{x}_k) \bar{\gamma}^2} \frac{\overline{P}_{\text{RZF}}^{(\mathcal{R})} \left(1 - \tau_k^2 + \tau_k^2 (1 + \bar{\gamma})^2\right) + \frac{\sigma^2}{l(\mathbf{x}_k)} (1 + \bar{\gamma})^2}{(1 - \tau_k^2)} \qquad (28)$$

*with $\tau_k = 0$ if $k \in \mathcal{S}_\mathcal{R}$.*

As it is seen, $\rho^\star$ does not depend on $\{\tau_k\}$ and it is basically in the same form of the perfect CSI case with the exception of the term $1 - c_S$ that accounts for the interference nulling towards the SCAs. Indeed, if no SCAs are active in the network, then $c_S = 0$ and $\rho^\star$ takes the same form in [17].

Since $\overline{P}_{\text{RZF}}^{(\mathcal{R})}$ must be positive and finite, from Lemma 2 it is seen that the following condition must be satisfied:

$$\frac{1}{K} \sum_{k \in \mathcal{M}_\mathcal{R}} \gamma_k \frac{\tau_k^2}{1 - \tau_k^2} < \frac{\rho^\star}{c} \bar{\gamma} \qquad (29)$$

from which setting $\tau_k = \tau$ for any $k \in \mathcal{M}_\mathcal{R}$ one gets

$$\tau < \tau_{\text{RZF}}^{(\max)} = \left(1 + \frac{c}{\rho^\star} \frac{\bar{\gamma}^{(\mathcal{M})}}{\bar{\gamma}}\right)^{-1/2} \qquad (30)$$

where $\bar{\gamma}^{(\mathcal{M})}$ is given by (17).

### B. Zero Forcing

Setting $\mathbf{\Lambda} = \mathbf{I}_K$ into (21) yields $\mathbf{F} = (\widehat{\mathbf{U}}\widehat{\mathbf{U}}^\dagger + N\rho \mathbf{I}_N)^{-1}\widehat{\mathbf{U}}$ from which using the Woodbury matrix identity and imposing $\rho = 0$ the ZF precoder $\mathbf{V}_{\text{ZF}} = \mathbf{T}^{(\mathcal{S}_\mathcal{B})}\widehat{\mathbf{U}}(\widehat{\mathbf{U}}^\dagger\widehat{\mathbf{U}})^{-1}$ easily follows or, equivalently,

$$\mathbf{V}_{\text{ZF}} = \mathbf{T}^{(\mathcal{S}_\mathcal{B})}\widehat{\mathbf{H}}(\widehat{\mathbf{H}}^\dagger \mathbf{T}^{(\mathcal{S}_\mathcal{B})}\widehat{\mathbf{H}})^{-1}. \qquad (31)$$

**Lemma 3.** *If ZF is used and $K, N \to \infty$ with $c + c_S \in (0, 1)$, then $P_{\text{ZF}}^{(\mathcal{R})} - \overline{P}_{\text{ZF}}^{(\mathcal{R})} \xrightarrow{a.s.} 0$ with*

$$\overline{P}_{\text{ZF}}^{(\mathcal{R})} = c\sigma^2 \frac{A}{1 - c_S - c(B+1)}. \qquad (32)$$

*Also, $p_k^{(\mathcal{R})} - \overline{p}_k^{(\mathcal{R})} \xrightarrow{a.s.} 0$ with*

$$\overline{p}_k^{(\mathcal{R})} = \frac{\gamma_k}{1 - \tau_k^2} \left(\sigma^2 + \tau_k^2 l(\mathbf{x}_k) \overline{P}_{\text{ZF}}^{(\mathcal{R})}\right) \qquad (33)$$

*with $\tau_k = 0$ if $k \in \mathcal{S}_\mathcal{R}$.*

*Proof:* The proof follows the same steps of that for Lemma 2 and thus is omitted for space limitations. ∎

From (32), it follows that the following condition must be satisfied: $1 - c_S - c(B+1) > 0$ or, equivalently,

$$\frac{1}{K} \sum_{k \in \mathcal{M}_\mathcal{R}}^{K} \gamma_k \frac{\tau_k^2}{1 - \tau_k^2} < \frac{1 - c - c_S}{c}. \qquad (34)$$

If $\tau_k = \tau$ for any $k$, then we have that

$$\tau < \tau_{\text{ZF}}^{(\max)} = \left(1 + \bar{\gamma}^{(\mathcal{M})} \frac{c}{1 - c - c_S}\right)^{-1/2}. \qquad (35)$$

From (30), it is seen that $\tau_{\text{RZF}}^{(\max)}$ is always larger than $\tau_{\text{ZF}}^{(\max)}$, meaning that RZF is more robust than ZF to imperfect CSI of MUEs. Observe that the same condition as in (35) must be fulfilled in the UL (see (16) in Section III).

## V. NUMERICAL RESULTS

Monte-Carlo simulations are now used to show that the above asymptotic characterization provides an effective means to evaluate the performance of a network with finite size. The results are obtained for 1000 different channel realizations and UE distributions. We assume that the BS is equipped with $N = 128$ antennas and covers a square area centered at the BS with side length 500m over which 16 SCAs are distributed on a regular grid with an inter-site distance of 125 m. We assume that 128 MUEs are active in the cell and that a single SUE is uniformly distributed within a disc of radius 35 m around each SCA. The SUEs are associated with the closest SCA while the MUEs are associated with the BS. Accordingly, the two sets $\mathcal{R}$ and $\mathcal{B}$ count 64 MUEs and 8 SCAs with 8 MUEs in the proximity of each SCA. A random snapshot of the network is depicted in Fig. 4. We assume that the UL and DL wireless backhaul rates of SCAs $r_s^{(\mathcal{S}_\mathcal{B},\text{dl})}$ and $r_s^{(\mathcal{S}_\mathcal{B},\text{ul})}$ are equal and fixed to 3 bit/s/Hz. The pathloss function $l(\mathbf{x})$ is [22]

$$l(\mathbf{x}) = 2L_{\bar{x}} \left(1 + \frac{\|\mathbf{x}\|^\beta}{\bar{x}^\beta}\right)^{-1} \qquad (36)$$

where $\beta > 2$ is the pathloss exponent, $\bar{x} > 0$ is some cut-off parameter and $L_{\bar{x}}$ is a constant that regulates the attenuation at distance $\bar{x}$. We assume that $\beta = 3.5$ and $L_{\bar{x}} = -86.5$ dB. The latter is such that for $f_c = 2.4$ GHz the attenuation at $\bar{x}$ is the same as that in the cellular model analyzed in [26].

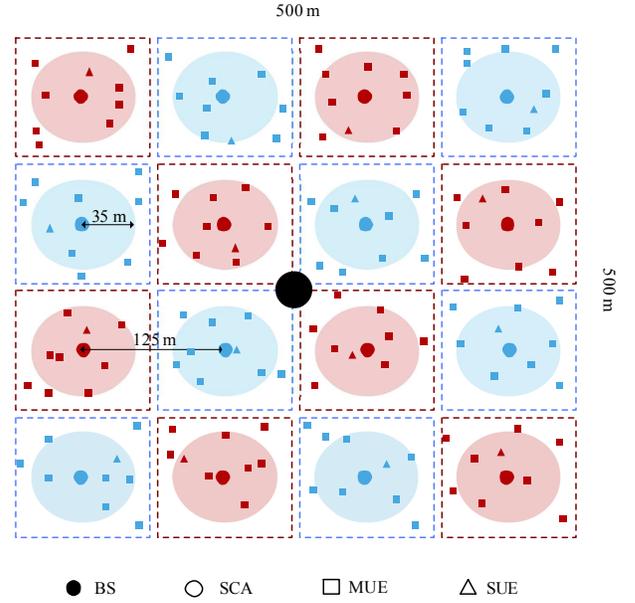

Fig. 4. A snapshot of the UE distribution in the simulated network wherein the number of SCAs is 16 and the number of MUEs is 128. The latter are distributed such that 8 of them are in the proximity of the coverage are of a given SCA. A single SUE is active for each SCA.

Although in TDD systems the effective values of $\{\tau_k\}$ are expected to be different between UL and DL (since the channels are estimated in the UL and then used in DL), the same values of $\{\tau_k\}$ are used for both links in all subsequent simulations. In particular, we assume $\tau_k = \tau \ \forall k$ and let $\tau^2 = \underline{\tau}^2 + \varsigma^2$ where $\underline{\tau}^2$ is basically modelled as a constant term (that basically accounts for pilot contamination, noisy measurements and other sources of estimation errors) while $\varsigma$ accounts for estimation errors induced by mobility. Following [12], [16], we set $\underline{\tau}^2 = 0.08$ while we compute $\varsigma^2$ as follows [21]

$$\varsigma^2 = 1 - J_0^2\left(2\pi \frac{v}{\lambda}\zeta\right) \qquad (37)$$

where $J_0(\cdot)$ denotes the 0-th order Bessel function of the first kind, $v$ is the velocity (in m/s) of MUEs, $\lambda$ is the carrier wavelength (in meter) and $\zeta$ is the UL or DL slot duration

TABLE I
GENERAL SYSTEM PARAMETERS

| Parameter | Value | Parameter | Value |
|---|---|---|---|
| Bandwidth | $W = 10$ MHz | Total number of SCAs | 16 |
| Noise power | $W\sigma^2 = -104$ dBm | Small-cell radius | $R = 35$ m |
| Macro-cell side length | 500 m | Inter-side distance of SCAs | $\Delta = 125$ m |
| Cut-off parameter | $\bar{x} = 25$ m | Pathloss coefficient | $\beta = 3.5$ |
| Carrier frequency | $f_c = 2.4$ GHz | Average pathloss attenuation at $\bar{x}$ | $L_{\bar{x}} = -86.5$ dB |
| Number of BS antennas | $N = 128$ | Imperfect CSI for MUEs | $\tau^2 = 0, 0.1$ and $0.3$ |
| Total number of MUEs | 128 | Wireless Backhaul Requirements for SCAs | $r_s = 3$ bit/s/Hz |

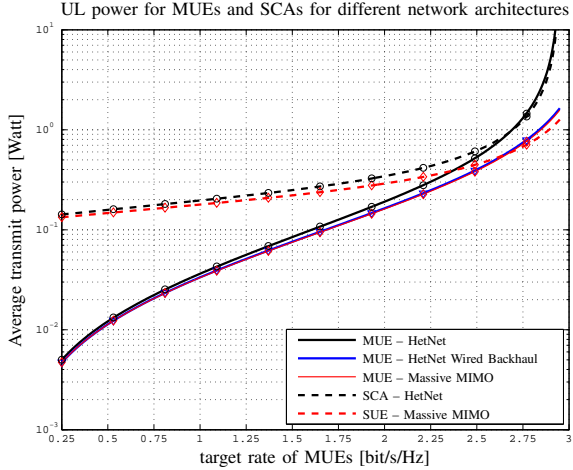

Fig. 5. Average UL transmit power for MUEs and SCAs as a function of MUE rates for different network architectures when $\tau^2 = 0.1$ and the wireless backhaul traffic is fixed to 3 bit/s/Hz.

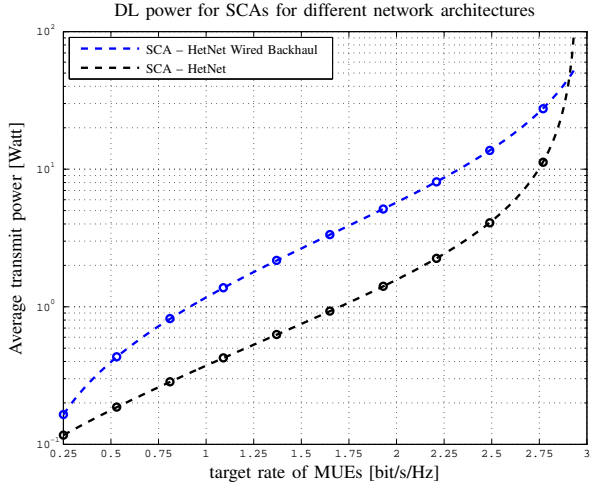

Fig. 6. Average DL transmit power for SCAs towards their respective SUEs as a function of MUE rates for the HetNet and HetNet with wired Backhaul architectures when $\tau^2 = 0.1$ and the wireless backhaul traffic is fixed to 3 bit/s/Hz.

(in seconds). Since $\lambda = 0.125$ m, setting $\zeta = 1$ ms and $v = 15$ or 50 km/h yields $\tau^2 = 0.1$ and 0.3, respectively. The parameter setting is summarized in Table I for simplicity. Comparisons are made with the two alternative protocols and network configurations mentioned in Remark 1. In particular, we consider a HetNet in which the SCAs use a wired backhaul infrastructure for data traffic and a massive MIMO system in which all UEs (MUEs and SUEs) are served by the macro BS. Observe that marker symbols correspond to Monte Carlo simulations while solid lines are based on the analytic results.

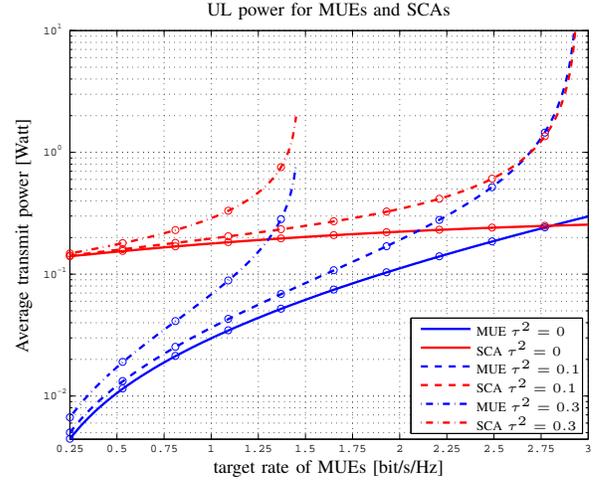

Fig. 7. Average UL transmit power for MUEs and SCAs for the proposed HetNet architecture as a function of MUE rates for different values of $\tau^2$. The wireless backhaul traffic is set to 3 bit/s/Hz.

Fig. 5 depicts the average transmit power for MUEs and SCAs over the cell as a function of the requested rate of MUEs when $\tau^2 = 0.1$. Despite the fact that the power of SCAs in (7) does not depend explicitly on the MUE rates $r_k^{(\mathcal{R},\text{ul})}$ for $k \in \mathcal{M}_\mathcal{R}$, a mild dependence on MUE requirements is shown in the results of Fig. 5. This is due to the fact that increased target rates for the MUEs result to increased overall interference in the system. A similar behaviour is observed in Fig. 6 for the DL power of SCAs. From the results of Fig. 5, it also follows that the average UL power of MUEs in the proposed HetNet is essentially the same of a HetNet with wired backhaul even though a wireless backhaul traffic of 3 bit/s/Hz is provided. In addition, it shows that the uplink transmit powers of SCAs are quite close to those of SUEs in the Massive MIMO case. On the contrary, Fig. 6 shows a significant power reduction for SCAs in DL mode. This is because the reception of associated SUEs is properly shielded from nearby MUEs in the proposed network architecture.

Figs. 7 and 8 provide insights on the effect of channel uncertainty on power consumption. Here again the target rates for the SCAs are fixed to 3 bits/s/Hz. Clearly, $\tau^2 = 0$ corresponds to the perfect CSI case. It can be seen that for $\tau^2 = 0.3$ (corresponding to a velocity of 50 km/h) the system becomes infeasible if the MUE target rates go beyond a certain level given approximately by 1.5 bit/s/Hz (as obtained through (16)). As seen, the power rapidly increases within a relatively narrow window beyond those rate values, thereby allowing the system to operate at relatively low powers up to close the critical points.

Figs. 9 and 10 illustrate the average DL transmit power of the BS when RZF and ZF are employed with $\tau^2 = 0.1$. As expected, RZF provides a substantial power reduction with respect to ZF. In particular, we observe that for a target rate of 2 bit/s/Hz only 0.1 W are required at the BS to serve (in DL) all MUEs and SCAs. Compared to the massive MIMO system, a marginal increase of power is required by the proposed HetNet. However, this is achieved with the benefit of a substantial power saving at the SUEs. Indeed, numerical results reveal that for a target ergodic rate of 3 bit/s/Hz, the required power for a SUE is 0.85 mW in the HetNet case while it is 0.22 W for a massive MIMO system. Observe that this power saving at SUE is of paramount importance as it allows to prolong the lifetime of batteries. Compared to an HetNet with wired backhaul, the proposed architecture achieves a substantial power saving especially for large target rates. As for the UL, this is because the interference from SUEs is properly shielded.

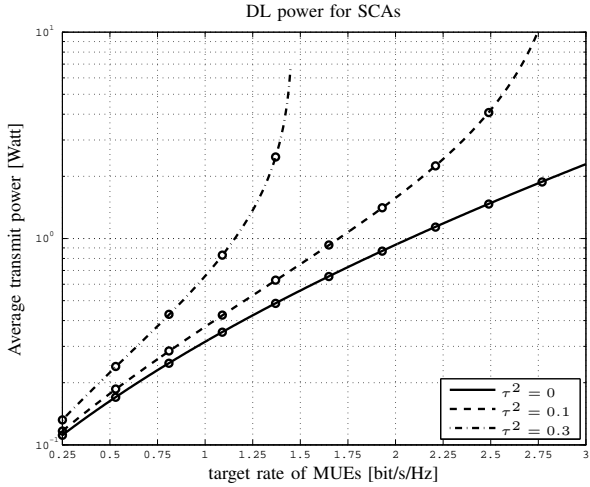

Fig. 8. Average DL transmit power for SCAs in the proposed HetNet architecture as a function of MUE rates for different values of $\tau^2$. The wireless backhaul traffic is set to 3 bit/s/Hz.

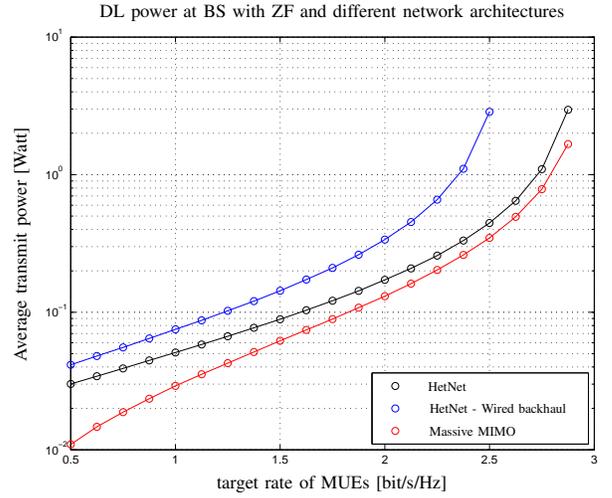

Fig. 10. Average DL transmit power at the BS when ZF is employed with $\tau^2 = 0.1$ and wireless backhaul 3 bit/s/Hz. Comparisons are made with a HetNet with wired backhaul and a single-tier massive MIMO systems operating according to the transmission protocols of Fig. 3.

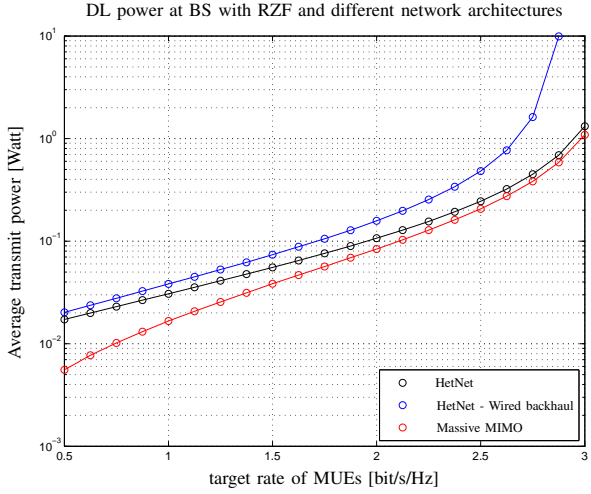

Fig. 9. Average DL transmit power at the BS when RZF is employed with $\tau^2 = 0.1$ and wireless backhaul 3 bit/s/Hz. Comparisons are made with a HetNet with wired backhaul and a single-tier massive MIMO systems operating according to the transmission protocols of Fig. 3.

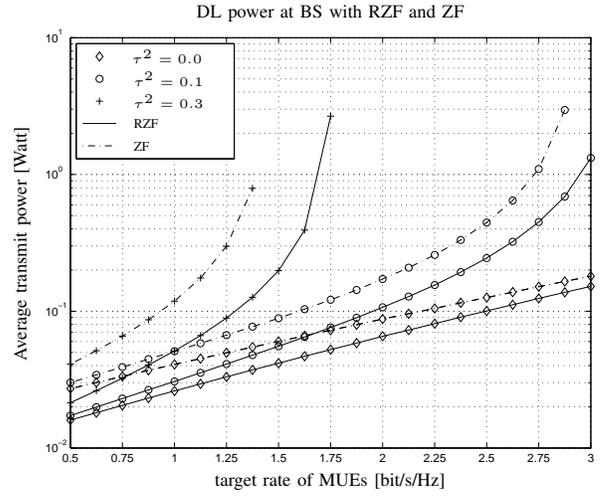

Fig. 11. Average DL transmit power at the BS when RZF and ZF are employed with different values of $\tau^2$ and wireless backhaul 3 bit/s/Hz.

Fig. 11 reports the average DL transmit power at the BS of RZF and ZF when $\tau^2 = 0, 0.1$ and 0.3. As seen, for $\tau^2 =$ 0.3 the power required by both precoding techniques diverge when the MUE target rate increases. As expected, RZF is more robust than ZF to imperfect CSI and can handle rates up to 1.75 bit/s/Hz.

To facilitate comparisons and highlight the potential gains of the proposed HetNet, in Table II we report the UL and DL power consumptions in the network. In particular, we consider the case of a MUE target rate of 1.5 bit/s/Hz and a velocity of 15 km/h (i.e., $\tau^2 = 0.1$). From the simulations above, it follows that the average UL power of each MUE is 83 mWatt. Taking into account that the bandwidth is 10 MHz and the number of MUEs for each frequency band is 64, this corresponds to an aggregate area throughput of 3.84 Gb/s/km$^2$. An additional throughput of 0.96 Gbit/s/km$^2$ comes from the 8 SUEs transmitting to the BS through the SCAs at target rate of 3 bit/s/Hz. This occurs at the cost of 0.25 Watt for each SCA while the power consumed by a SUE is relatively small and given by 0.85 mWatt. In the DL, the same throughput as for the UL is achieved consuming only 55 mWatt at the BS and 0.75

TABLE II
POWER CONSUMPTION IN WATT OF THE DIFFERENT ARCHITECTURES FOR A MUE TARGET RATE OF 1.5 BIT/S/HZ WITH $\tau^2 = 0.1$

| (Watt) | HetNet | HetNet - Wired Backhaul [9] | Massive MIMO |
|---|---|---|---|
| MUE (UL) | 0.083 | 0.075 | 0.075 |
| SCA (UL) | 0.25 | -- | -- |
| SUE (UL) | $8.5 \cdot 10^{-4}$ | $8.5 \cdot 10^{-4}$ | 0.22 |
| BS (DL) | 0.055 | 0.074 | 0.038 |
| SCA (DL) | 0.75 | 2.69 | -- |

Watt at the SCA. Putting everything together, it follows that a total aggregate area throughput of 4.8 Gb/s/km² is achieved with a total power consumption of only 5.52 Watt in the UL and 6.05 Watt in the DL, thereby showing the potential gains of the proposed HetNet. Note that the total power consumption of the massive MIMO network is 8.32 Watt in the UL while it is only 0.038 Watt in the DL. As mentioned above, however, this is achieved at the price of a large increase of the transmit power at SUEs (up to 0.22 W) compared to the proposed HetNet (only 0.85 mWatt).

## VI. DISCUSSIONS AND PERSPECTIVES

In this section, we discuss the impact of mobility along with some other practical aspects of the proposed HetNet.

### A. Impact of mobility

In [12], the authors show that if the network sum rate is considered then low and high mobility MUEs can coexist and be served simultaneously. This is because the imperfect CSI of each given MUE has detrimental effects only on its own achievable rate while it has no impact on the performance of the others. This is in sharp contrast to the results obtained in this work where we have shown that the UL and DL transmit powers for meeting target rates depend heavily on the mobility of each MUE. In particular, a single MUE with high mobility and rate requirements might largely increase the required powers. This calls for alternative solutions.

The simplest one would be to lower the target rate (and thus the corresponding SINR) for the MUEs with large channel estimation errors such that, for example, in UL the condition for $\delta \to 0$ in (15) is satisfied. An alternative solution for DL mode might consist in dividing the MUEs in two sets $\mathcal{A}_M$ and $\mathcal{A}_H$ (with $\mathcal{A}_M \cap \mathcal{A}_H = \emptyset$) characterized by medium and high mobility, respectively. The MUEs in $\mathcal{A}_M$ are served simultaneously while those in $\mathcal{A}_H$ are served one at a time using space-time coding (STC) techniques (that do not require any CSI at the BS). Consider for example the frequency-time slot $(W_1, T_2)$ for which $\mathcal{M}_\mathcal{R} = \mathcal{A}_M \cup \mathcal{A}_H$. Let $K = |\mathcal{A}_M| + |\mathcal{S}_\mathcal{R}|$, $K_{STC} = |\mathcal{A}_H|$ and $S = |\mathcal{S}_\mathcal{B}|$. The BS would first simultaneously transmit to the $K$ MUEs and SCAs in $\mathcal{R} = \mathcal{A}_M \cup \mathcal{S}_\mathcal{R}$ while removing the interference towards to the $S$ SCAs in $\mathcal{S}_\mathcal{B}$. Then, it would serve the $K_{STC}$ MUEs in $\mathcal{A}_H$ (one at a time) while nulling the interference to $\mathcal{S}_\mathcal{B}$. In these circumstances, the signal transmitted to MUE $k$ in $\mathcal{A}_H$ takes the form $\mathbf{x}_k = \mathbf{T}^{(\mathcal{S}_\mathcal{B})} \mathbf{s}_k$ with $\mathbf{s}_k$ being such that $\mathbb{E}\{\mathbf{s}_k \mathbf{s}_k^\dagger\} = p_k^{(\mathcal{A}_H, \text{dl})}/N \mathbf{I}_N$ (corresponding to uniform STC). As a consequence, the deterministic equivalent of the DL SINR of MUE $k$ in $\mathcal{A}_H$ is found to be:

$$\text{SINR}_k^{(\mathcal{A}_H, \text{dl})} - (1 - c_S) \frac{p_k^{(\mathcal{A}_H, \text{dl})} l(\mathbf{x}_k)}{\sigma^2} \xrightarrow{a.s.} 0 \quad (38)$$

from which (using dominated convergence arguments and continuous mapping theorem) it follows that $p_k^{(\mathcal{A}_H, \text{dl})} - \overline{p}_k^{(\mathcal{A}_H, \text{dl})} \xrightarrow{a.s.} 0$ with $\overline{p}_k^{(\mathcal{A}_H, \text{dl})} = \frac{1}{1 - c_S} \frac{\gamma_k \sigma^2}{l(\mathbf{x}_k)}$.

Let $T_{STC}$ be the time required to serve the $K_{STC}$ MUEs in $\mathcal{A}_H$ and call $T_{LP} = T_2 - T_{STC}$ where LP stands for linear precoding and $T_2$ is defined in Fig. 2. Accordingly, the average spectral efficiency $R_{\text{AVG}}$ (in bit/s/Hz) of the network over $T_2 = T_{LP} + T_{STC}$ is

$$R_{\text{AVG}} = \frac{T_{LP}}{T_2} \sum_{k=1}^{K} \log_2(1 + \gamma_k) + \frac{T_{STC}}{T_2 K_{STC}} \sum_{k=K+1}^{K+K_{STC}} \log_2(1 + \gamma_k)$$

and the corresponding energy consumption is obtained as

$$\overline{E}_{T_2} = \frac{c \sigma^2 A T_{LP}}{1 - c_S - c(B+1)} + T_{STC} \sum_{k=K+1}^{K+K_{STC}} \frac{1}{1 - c_S} \frac{\gamma_k \sigma^2}{l(\mathbf{x}_k)}.$$

As seen, the rate of MUEs served by STC is reduced by a factor $1/K_{STC}$ compared to the other ones if $T_{STC} \approx T_{LP}$. On the other hand, if $T_{STC} \approx K_{STC} T_{LP}$ then the spectral efficiencies are comparable, but the energy consumption increases substantially.

### B. Tradeoff between proximity effect and density of users

A close inspection of (13) reveals that the interference term $I_K = \sum_{k=1}^{K} p_k^{(\mathcal{R})} l(\mathbf{x}_{sk})$ for SUEs in $\mathcal{S}_\mathcal{B}$ increases with $K$. This means that, although spatially separated, the interference from UL signals in $\mathcal{R}$ might be large (due to the possibly large values of $K$). For example, for the setting of Table II the average interference level (normalized to the noise power) is numerically found to be $1.3 \times 10^3$. Although pretty significant, this interference level does not prevent the network to properly operate since SUEs experience (on average) good SINRs due to their proximity to the SCAs. Clearly, this is a consequence of the specific network under consideration and, in particular, it largely depends on the SCA radius $R$, the inter-SCA-location distance $\Delta$ and the MUE density. All these parameters play a key role in determining the SINR of SUEs. Unfortunately, a theoretical analysis revealing the interplay among all of them is a challenging problem. To partially address this issue, we resort to a kind of worst case scenario in which the MUEs transmit with constant power, i.e., $p_k^{(\mathcal{R})} = \overline{p}$, and are uniformly distributed with density $\alpha$ in an infinite cell with the only exception of a circle of radius $d = \Delta/2 - R$ around the SUE.[5] Under the above assumptions, we have that (details are omitted

---

[5]From Fig. 4, it follows that $d$ is actually the minimum distance $\|\mathbf{x}_{sk}\|$ between any SUE-MUE pair since it corresponds to the extreme case of the SUE being at distance $R$ from the SCA and the MUE being at the closest point to it.

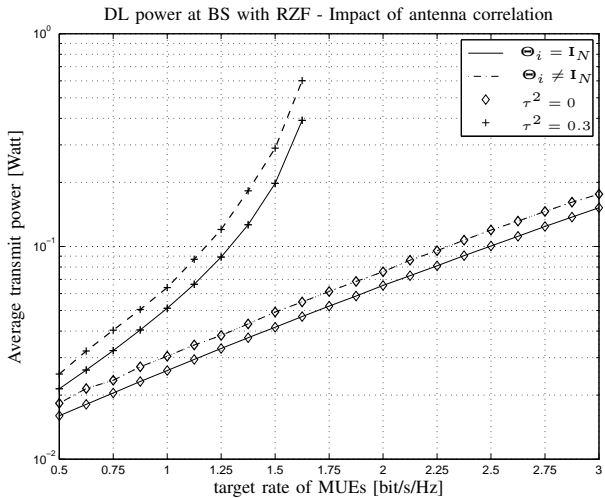

Fig. 12. Average DL transmit power at the BS when RZF is employed with different values of $\tau^2$ and wireless backhaul 3 bit/s/Hz. Comparisons are made with the case in which BS antennas are correlated.

for space limitations):

$$\mathbb{E}\{I_K\} = \frac{4\pi\alpha\overline{p}L_{\bar{x}}}{\beta-2}\frac{\bar{x}^\beta}{d^{\beta-2}} F\left(1, 1-\frac{2}{\beta}; 2-\frac{2}{\beta}, -\left(\frac{\bar{x}}{d}\right)^\beta\right) \quad (39)$$

where $F(\cdot)$ is the hypergeometric function [27]. Plugging the system parameters of Table I into the above equation and setting $\overline{p} = 83$ mWatt (as specified in Table II) yields $\mathbb{E}\{I_K\}/\sigma^2 \approx 5.5 \times 10^3$, which is of the same order of the one evaluated numerically and given by $1.3 \times 10^3$. This validates the above analysis and makes (39) accurate enough although derived under simplifying assumptions. Interestingly, if $d \gg \bar{x}$ then (39) reduces to $\mathbb{E}\{I_K\} \approx \frac{4\pi\alpha\overline{p}L_{\bar{x}}}{\beta-2}\frac{\bar{x}^\beta}{d^{\beta-2}}$ from which the value of $d$ required to keep the interference level at a prescribed value (for a given MUE density) can be easily obtained. Moreover, observing that $l(d) \approx \bar{x}^\beta/d^\beta$ for $d \gg \bar{x}$ we obtain $\mathbb{E}\{I_K\} \approx \frac{2\pi}{\beta-2}\alpha d^2 \overline{p} l(d)$, which has the following very intuitive explanation: The average interference is (up to a proportionality factor $2\pi/(\beta-2)$) only due to the $\alpha d^2$ MUEs located at a distance of order $d$ from the SUE. This provides a simple way to evaluate the tradeoff between the density of users and the minimum separation between MUEs and SUEs.

Observe that the above analysis is valid only when the MUEs are uniformly distributed. If this assumption does not hold true, the problem is much more involved since the location itself of SCAs should be also optimized taking into account the spatially varying MUE densities. However, such a case is beyond the scope of this work and left for future research.

### C. Impact of correlation at BS antennas

If the BS antennas are correlated, then the channel vector of the $i$th device in $\mathcal{A}$ is modelled as $\mathbf{h}_i^{(\mathcal{A})} = \sqrt{Nl(\mathbf{x}_i)}\boldsymbol{\Theta}_i^{1/2}\mathbf{w}_i$ where $\boldsymbol{\Theta}_i$ denotes the $i$th channel correlation at the BS [16]. As for positions $\mathbf{x}_i$, the matrices $\boldsymbol{\Theta}_i$ are usually assumed to change slowly compared to the channel coherence time and thus are supposed to be perfectly known at the BS [16], [28]. As a consequence, $\widehat{\mathbf{h}}_i^{(\mathcal{A})}$ can be reasonably modelled as $\widehat{\mathbf{h}}_i^{(\mathcal{A})} = \sqrt{Nl(\mathbf{x}_i)}\boldsymbol{\Theta}_i^{1/2}\widehat{\mathbf{z}}_i$ with $\widehat{\mathbf{z}}_i$ still given by (2). Assume that RZF is employed. Then, $\mathbf{V}_{\text{RZF}} = \mathbf{T}^{(\mathcal{S_B})}\mathbf{F}$ with

$$\mathbf{F} = \left(N\sum_{i=1}^{K}\tilde{\boldsymbol{\Theta}}_i^{1/2}\widehat{\mathbf{z}}_i\widehat{\mathbf{z}}_i^\dagger\tilde{\boldsymbol{\Theta}}_i^{1/2} + N\rho\mathbf{I}_N\right)^{-1}\widehat{\mathbf{U}} \quad (40)$$

where $\widehat{\mathbf{U}} = \mathbf{T}^{(\mathcal{S_B})}\widehat{\mathbf{H}}$ and $\tilde{\boldsymbol{\Theta}}_i^{1/2} = \mathbf{T}^{(\mathcal{S_B})}\boldsymbol{\Theta}_i^{1/2}$. Following the same steps in [16], [28], one can in principle compute the deterministic equivalents of $\{p_k^{(\mathcal{R})}\}$ and $P_{\text{RZF}}^{(\mathcal{R})}$ for $\boldsymbol{\Theta}_i \neq \mathbf{I}_N$. Although possible (not shown for space limitations), this ends up to compute the fixed point of a set of equations and to evaluate the inverse of $K \times K$ matrix. All this is not only much more involved than the case $\boldsymbol{\Theta}_i = \mathbf{I}_N$ but it is also less instrumental to get insights into the structure of the asymptotic transmit powers and into the interplay among the different parameters (such as imperfect CSI factors $\{\tau_k\}$, SINR constraints $\{\gamma_k\}$ and number of MUEs and SCAs). In addition, the optimal regularization parameter $\rho$ can only be found through a numerical optimization procedure.

Fig. 12 reports the DL transmit power when the BS is equipped with RZF and antennas are correlated. Following [28], the entries of $\boldsymbol{\Theta}_k$ for $k = 1, 2, \ldots, K$ are computed as

$$[\boldsymbol{\Theta}_k]_{i,\ell} = \frac{1}{\Delta_\varphi}\int_{\theta_k-\Delta_\varphi/2}^{\theta_k+\Delta_\varphi/2} e^{\mathrm{i}\pi\cos(\varphi)}\partial\varphi \quad (41)$$

where $\Delta_\varphi$ is the angular spread and $\theta_k$ is the directional of departure of the $k$th signal. We set $\Delta_\varphi = \pi/12$ and assume that $\theta_k$ are uniformly distributed in $[0, 2\pi)$. Only a marginal difference is observed in terms of required power between the two cases. Moreover, imperfect CSI has the same impact in both cases. A similar behaviour is obtained for larger values of $\Delta_\varphi$ up to $\Delta_\varphi = \pi/6$.

### D. Dynamic UL-DL TDD

The transmission protocol of Fig. 2 relies on the assumption that transmissions across tiers are perfectly synchronized. However, the synchronous operation with a common UL and DL configuration in multiple cells may not match the instantaneous traffic situation in a particular cell. The amount of traffic for DL and UL may vary significantly with time and between cells. This calls for the adoption of a dynamic UL-DL configuration [24]. Henceforth, we discuss some practical implications of dynamic TDD for the proposed network architecture. Consider for example the frequency band $W_1$ in Fig. 2 and assume that the SCAs in $\mathcal{S_R}$ are not aligned with the UL and DL transmissions of MUEs in $\mathcal{M_R}$. If for example the UL phase BS $\leftarrow$ SCA is shorter than BS $\leftarrow$ MUE, then the subsequent DL phase BS $\rightarrow$ SCA would partially overlap with BS $\leftarrow$ MUE. A similar situation would occur if the SCAs are in the UL for a longer time interval. In both cases, the adoption of a dynamic TDD protocol at the SCAs would require a full duplex BS. On the other hand, if the SCAs in $\mathcal{S_B}$ are not aligned with the MUEs in $\mathcal{M_R}$, the following two situations might occur. If SCA $\rightarrow$ SUE is longer than BS $\leftarrow$ MUE, then the linear precoder at the BS must be designed so as to also mitigate interference towards the SUEs in $\mathcal{S_B}$. If SCA $\rightarrow$ SUE is shorter than BS $\leftarrow$ MUE, the SCAs in $\mathcal{S}_\mathcal{B}$ are affected by the interference due to UL transmissions in $\mathcal{M}_\mathcal{R}$ and $\mathcal{S}_\mathcal{R}$. The effect of this interference would be the same of that evaluated in Section III. In summary, the proposed network architecture and transmission protocol allow dynamic TDD transmissions within the small-cell tier (from SCAs to SUEs) while a full duplex BS would be required to handle asynchronous transmissions at the macro-tier level.

## VII. CONCLUSIONS

This work has focused on the power consumption in the UL and DL of a HetNet in which a massive MIMO macro tier (serving medium-to-high mobility UEs) is overlaid with a dense tier of SCAs using a wireless backhaul for traffic. A reverse (inter-tier and intra-tier) TDD protocol has been proposed to let the BS simultaneously handle the traffic of macro UEs and SCAs without causing much interference to the overlaid tier. Linear processing has been used at the BS for data recovery and transmission while satisfying rate requirements and mitigating interference. In particular, we have considered an MMSE receiver and a concatenated linear precoding technique based on ZF and RZF. The analysis has been conducted in the asymptotic regime where the number of BS antennas and network size grow large with fixed ratio. Results from random matrix theory have been used to derive closed-form expressions for the transmit powers and beamforming vectors as well as to investigate the impact of imperfect CSI on the power consumption. It turns out that for a given set of target rates there is a critical value of imperfect CSI beyond which the power of all transmitters rapidly increases (and eventually diverges). However, analytical and numerical results have shown that when such critical values are not met the proposed architecture allows to achieve an aggregate area throughput on the order of 4.8 Gb/s/km$^2$ in UL and DL on a 10 MHz band with a very limited amount of power on the order of 6 Watt in both UL and DL.

An important follow-up of this work could be the development of scheduling algorithms for serving MUEs characterized by very high mobility without lowering the served rates. The extension of the analysis to a multi-cell network in which multiple BSs (with limited cooperation) are active is also very much interesting. This could be addressed using the results in [29]–[31]. The large system analysis used throughout this work could in principle be also used for other macro-diversity studies such as those in [32]. An interesting problem is also to develop network architectures able to exploit the gains of massive MIMO when the macro tier level operates according to a frequency division duplexing (FDD) system.

## APPENDIX

*1) Proof of Lemma 1:* The proof builds upon applying the asymptotic results shown in Appendix II of [16] under the assumption that the correlation matrix of $\mathbf{h}_k$ is given by $l(\mathbf{x}_k)\mathbf{I}_N$. More precisely, the deterministic equivalent of $\mathbf{g}_k^\dagger \mathbf{h}_k$ follows directly from [16] by taking into account that $\mathbf{G}$ in (4) includes the powers $p_k^{(\mathcal{R},\mathrm{ul})}$. Omitting the mathematical details for space limitations, we have that

$$\mathbf{g}_k^\dagger \mathbf{h}_k - \sqrt{1-\tau_k^2}\frac{\xi l(\mathbf{x}_k)}{1+\xi l(\mathbf{x}_k) p_k^{(\mathcal{R},\mathrm{ul})}} \xrightarrow{a.s.} 0 \qquad (44)$$

where $\xi$ is given by (8). The asymptotic expression of $\mathbf{g}_k^\dagger \mathbf{g}_k$ is found to be [16]

$$\mathbf{g}_k^\dagger \mathbf{g}_k + \frac{l(\mathbf{x}_k)\xi'}{\left(1+\xi l(\mathbf{x}_k)p_k^{(\mathcal{R},\mathrm{ul})}\right)^2} \xrightarrow{a.s.} 0 \qquad (45)$$

and is obtained by simply noting that $\mathbf{g}_k^\dagger \mathbf{g}_k = -(\widehat{\mathbf{h}}_k^\dagger \mathbf{g}_k)'$ where $(\cdot)'$ denotes the derivative with respect to $\sigma^2$. To compute the deterministic equivalent of $|\mathbf{g}_k^\dagger \mathbf{h}_i|^2$, we apply the Woodbury identity twice and use the same arguments as in [16] to obtain

$$-\left(\tau_i^2 + \frac{1-\tau_i^2}{(1+\xi l(\mathbf{x}_i)p_i^{(\mathcal{R})})^2}\right)\frac{l(\mathbf{x}_i)l(\mathbf{x}_k)\xi'}{(1+\xi l(\mathbf{x}_k)p_k^{(\mathcal{R})})^2}. \qquad (46)$$

The deterministic equivalent of $|\mathbf{g}_k^\dagger \mathbf{h}_s^{(\mathcal{S}_\mathcal{B})}|^2$ follows from the above results by recalling that $\tau_i = 0$ for SCA channels. Plugging everything together leads to (42) from which imposing $\mathrm{SINR}_k^{(\mathcal{R})} = \gamma_k$ the result follows using simple calculus.

*2) Proof of Lemma 2:* The proof follows the same steps of [16] with the only exception that the projection matrix $\mathbf{T}^{(\mathcal{S}_\mathcal{B})}$ must be included in the analysis. Note that $\widehat{\mathbf{U}}\mathbf{\Lambda}^{-1}\widehat{\mathbf{U}}^\dagger = \sum_{k=1}^K \mathbf{T}^{(\mathcal{S}_\mathcal{B})}\widehat{\mathbf{z}}_k \widehat{\mathbf{z}}_k^\dagger \mathbf{T}^{(\mathcal{S}_\mathcal{B})}$ such that

$$\mathbf{F} = \left(N\sum_{k=1}^K \mathbf{T}^{(\mathcal{S}_\mathcal{B})}\widehat{\mathbf{z}}_k \widehat{\mathbf{z}}_k^\dagger \mathbf{T}^{(\mathcal{S}_\mathcal{B})} + N\rho \mathbf{I}_N\right)^{-1}\mathbf{T}^{(\mathcal{S}_\mathcal{B})}\widehat{\mathbf{H}} \qquad (47)$$

which is exactly in the same form of the RZF precoder used in [16] when all the UEs have the same correlation matrix given by $\mathbf{T}^{(\mathcal{S}_\mathcal{B})}$ (using the notation of [16] this amounts to setting $\mathbf{\Theta}_k = \mathbf{T}^{(\mathcal{S}_\mathcal{B})}\ \forall k$). From the results of Theorem 2 in [16], it follows that if $\mathbf{T}^{(\mathcal{S}_\mathcal{B})}$ has uniformly bounded spectral norm on $N$ (i.e., $\lim_{N,K,S\to\infty}\sup\|\mathbf{T}^{(\mathcal{S}_\mathcal{B})}\| < \infty$) then

$$\mathrm{SINR}_k^{(\mathcal{R})} - \frac{\overline{p}_k^{(\mathcal{R})}(1-\tau_k^2)l(\mathbf{x}_k)\mu^2}{\overline{P}_{\mathrm{RZF}}^{(\mathcal{R})}\left(1-\tau_k^2+\tau_k^2(1+\mu)^2\right) + \frac{\sigma^2(1+\mu)^2}{l(\mathbf{x}_k)}} \xrightarrow{a.s.} 0$$

with $\tau_k = 0$ if $k \in \mathcal{M}_\mathcal{R}$ and $\mu$ being the solution of

$$\mu = \frac{1}{N}\mathrm{tr}\left(\mathbf{T}^{(\mathcal{S}_\mathcal{B})}\left(\mathbf{T}^{(\mathcal{S}_\mathcal{B})}\frac{c}{1+\mu} + \rho\mathbf{I}_N\right)^{-1}\right). \qquad (48)$$

Applying the Woodbury identity to $(\mathbf{T}^{(\mathcal{S}_\mathcal{B})}\frac{c}{1+\mu}+\rho\mathbf{I}_N)^{-1}$ with $\mathbf{T}^{(\mathcal{S}_\mathcal{B})}$ given in (20) and observing that $\mathrm{tr}(\mathbf{H}^{(\mathcal{S}_\mathcal{B})}(\mathbf{H}^{(\mathcal{S}_\mathcal{B})\dagger}\mathbf{H}^{(\mathcal{S}_\mathcal{B})})^{-1}\mathbf{H}^{(\mathcal{S}_\mathcal{B})\dagger}) = S$, then (48) becomes

$$\mu = (1-c_S)\left(\frac{c}{1+\mu}+\rho\right)^{-1}. \qquad (49)$$

The deterministic equivalent of $P_{\mathrm{RZF}}^{(\mathcal{R})}$ is found to be [16]

$$\overline{P}_{\mathrm{RZF}}^{(\mathcal{R})} = -\frac{c\mu'}{(1+\mu)^2}\frac{1}{K}\sum_{k=1}^K p_k l(\mathbf{x}_k) \qquad (50)$$

with $\mu' = -\frac{\mu(1+\mu)^2}{c+\rho(1+\mu)^2}$. Assume now that the power $\overline{p}_k^{(\mathcal{R})}$ is chosen such that $\mathrm{SINR}_k^{(\mathcal{R})}$ is equal to a specified $\gamma_k$ in the large system limit. Then, one gets

$$\overline{p}_k^{(\mathcal{R})} = \frac{\gamma_k}{l(\mathbf{x}_k)\mu^2}\frac{\overline{P}_{\mathrm{RZF}}^{(\mathcal{R})}\left(1-\tau_k^2+\tau_k^2(1+\mu)^2\right) + \frac{\sigma^2}{l(\mathbf{x}_k)}(1+\mu)^2}{1-\tau_k^2} \qquad (51)$$

$$\text{SINR}_k^{(\mathcal{R})} + \frac{\xi^2}{\xi'} \frac{(1-\tau_k^2)l(\mathbf{x}_k)p_k^{(\mathcal{R})}}{\frac{1}{N}\sum_{i=1}^{K} l(\mathbf{x}_i)p_i^{(\mathcal{R})}\left[\tau_i^2 + \frac{1-\tau_i^2}{\left(1+\xi l(\mathbf{x}_i)p_i^{(\mathcal{R})}\right)^2}\right] + \frac{1}{N}\sum_{s=1}^{S} \frac{l(\mathbf{x}_s)p_s^{(\mathcal{S}_{\mathcal{B}})}}{\left(1+\xi l(\mathbf{x}_s)p_s^{(\mathcal{S}_{\mathcal{B}})}\right)^2} + \sigma^2} \xrightarrow{a.s.} 0 \quad (42)$$

$$\overline{P}_{\text{RZF}}^{(\mathcal{R})} = -\frac{c\mu'}{(1+\mu)^2}\sum_{k=1}^{K}\frac{1}{K}\frac{\gamma_k}{(1-\tau_k^2)l(\mathbf{x}_k)\mu^2}\left(\overline{P}_{\text{RZF}}^{(\mathcal{R})}\left(1-\tau_k^2 + \tau_k^2(1+\mu)^2\right) + \frac{\sigma^2(1+\mu)^2}{l(\mathbf{x}_k)}\right) \quad (43)$$

with $\tau_k = 0$ if $k \in \mathcal{S}_\mathcal{R}$. Using the above result in (50) yields (43). Solving (43) with respect to $\overline{P}_{\text{RZF}}^{(\mathcal{R})}$ and taking the derivative with respect to $\rho$ yields (omitting the computations for simplicity)

$$\frac{\partial \overline{P}_{\text{RZF}}^{(\mathcal{R})}}{\partial \rho} = \frac{2c^2 A\sigma^2 (\bar{\gamma}-\mu)}{\left(\mu\left(c+\rho(1+\mu)^2\right) - c\left(\bar{\gamma}+B(1+\mu)^2\right)\right)^2}.$$

From the above result, it turns out that the minimum power is achieved when $\mu = \bar{\gamma}$. Plugging this result into (49) yields (27) from which the result in Lemma 2 follows from (50) and (51).